\def\isarxiv{1}

\def\paperTitle{Beyond Classical Attention: Quantum Attention for Scalable Computation}

\def\paperRTitle{\paperTitle} 

\def\paperAuthor{
Xuyang Guo\thanks{Guilin University of Electronic Technology.}
\and
Zhao Song\thanks{\texttt{magic.linuxkde@gmail.com}. Simons Institute for the Theory of Computing, UC Berkeley.}
\and
Xin Yang\thanks{The University of Washington.}
\and 
Ruizhe Zhang\thanks{Purdue University.}
}

\ifdefined\isarxiv
\documentclass[11pt]{article}
\usepackage[numbers]{natbib}
\else
\documentclass{article}

\fi

\ifdefined\isarxiv

\usepackage{amsmath}
\usepackage{amsthm}
\usepackage{amssymb}
\usepackage{algorithm}
\usepackage{subfig}
\usepackage{algpseudocode}
\usepackage{graphicx}
\usepackage{grffile}
\usepackage{wrapfig,epsfig}
\usepackage{url}
\usepackage{xcolor}
\usepackage{epstopdf}

\usepackage{bbm}
\usepackage{dsfont}

\else 

\usepackage{microtype}
\usepackage{graphicx}
\usepackage{subcaption}
\usepackage{booktabs} 

\usepackage{hyperref}

\usepackage{icml2026}

\usepackage{amsmath}
\usepackage{amssymb}
\usepackage{mathtools}
\usepackage{amsthm}

\fi

\ifdefined\isarxiv

\else
\usepackage{algpseudocode}
\fi
 
\allowdisplaybreaks

\ifdefined\isarxiv

\usepackage{tikz}
\usepackage{hyperref}  
\hypersetup{colorlinks=true,citecolor=blue,linkcolor=blue} 
\usetikzlibrary{arrows}
\usepackage[margin=1in]{geometry}

\else
\fi
 
\graphicspath{{./figs/}}

\theoremstyle{plain}
\newtheorem{theorem}{Theorem}[section]
\newtheorem{lemma}[theorem]{Lemma}
\newtheorem{definition}[theorem]{Definition}

\newtheorem{remark}[theorem]{Remark}

\newcommand{\wt}{\widetilde}

\newcommand{\R}{\mathbb{R}}

\newcommand{\ket}[1]{|#1\rangle}

\DeclareMathOperator{\poly}{poly}

\DeclareMathOperator{\diag}{diag}

\ifdefined\isarxiv

\else
\icmltitlerunning{\paperRTitle}

\fi

\begin{document}

\ifdefined\isarxiv

\date{}
\title{\paperTitle}
\author{\paperAuthor}

\else

\twocolumn[
  \icmltitle{\paperTitle}


  \icmlsetsymbol{equal}{*}

  \begin{icmlauthorlist}
    \icmlauthor{Firstname1 Lastname1}{equal,yyy}
    \icmlauthor{Firstname2 Lastname2}{equal,yyy,comp}
    \icmlauthor{Firstname3 Lastname3}{comp}
    \icmlauthor{Firstname4 Lastname4}{sch}
    \icmlauthor{Firstname5 Lastname5}{yyy}
    \icmlauthor{Firstname6 Lastname6}{sch,yyy,comp}
    \icmlauthor{Firstname7 Lastname7}{comp}
    \icmlauthor{Firstname8 Lastname8}{sch}
    \icmlauthor{Firstname8 Lastname8}{yyy,comp}
  \end{icmlauthorlist}

  \icmlaffiliation{yyy}{Department of XXX, University of YYY, Location, Country}
  \icmlaffiliation{comp}{Company Name, Location, Country}
  \icmlaffiliation{sch}{School of ZZZ, Institute of WWW, Location, Country}

  \icmlcorrespondingauthor{Firstname1 Lastname1}{first1.last1@xxx.edu}
  \icmlcorrespondingauthor{Firstname2 Lastname2}{first2.last2@www.uk}

  \icmlkeywords{Machine Learning, ICML}

  \vskip 0.3in
]

\printAffiliationsAndNotice{} 

\fi

\ifdefined\isarxiv
\begin{titlepage}
  \maketitle
  \begin{abstract}
    As large language models (LLMs) demonstrate outstanding performance across various tasks, attention-driven models have profoundly transformed the field of machine learning. Since attention computations account for the primary computational overhead in both model inference and training, efficiently computing attention matrices has become one of the core challenges in accelerating large language models. It is well-known that quantum machines possess computational advantages over classical machines, and the role of quantum computing in LLMs remains largely unexplored. In this work, we focus on leveraging the Grover search algorithm to efficiently compute a sparse attention matrix. Through comparisons with classical algorithms, we demonstrate that our method achieves quantum acceleration in polynomial time. Additionally, we observe that the generated quantum attention matrices naturally exhibit low-rank structures, providing further theoretical support for efficient modeling. Moreover, within the specific context of attention matrix computation, we conduct a systematic and detailed analysis of the error and time complexity of the proposed algorithm.

  \end{abstract}
  \thispagestyle{empty}
\end{titlepage}


\else

\begin{abstract}

\end{abstract}

\fi



\section{Introduction}
LLMs (Large Language Models)~\cite{s18,vkb23,kgw+23,emm+23} have gained significant attention from numerous researchers in recent years. The success of models like OPT~\cite{zrg+22}, PaLM~\cite{cnd+22}, GPT-3~\cite{bmr+20}, Transformer~\cite{vsp+17}, and BERT~\cite{dclt18} has showcased the immense potential of LLMs in various applications across different domains.

The impact of LLMs is far-reaching. They have revolutionized natural language processing tasks such as machine translation~\cite{hwl21}, sentiment analysis \cite{uas+20}, question answering \cite{bmr+20,o23}, text summarization, and more. LLMs excel at capturing intricate language patterns, understanding context, and generating coherent and contextually relevant text \cite{o23}.

The success of Transformer \cite{vsp+17} heavily relies on the multi-head attention algorithm, which plays a crucial role in the computation of LLM models. Models like GPT \cite{bmr+20,o23} have achieved remarkable success by employing a large number of parameters and leveraging vast amounts of data, which holds great potential for future advancements. However, this approach raises concerns about the computational running time, necessitating the development of more efficient algorithms \cite{bsz23,zhdk23,as23}. In this regard, the emergence of quantum algorithms \cite{g96,s99,hhl09} provides a new perspective for addressing this problem.

To demonstrate our design, we will begin by introducing the attention matrix and its classical computation. 
The attention matrix is a square matrix whose rows and columns are indexed by tokens (i.e., words), and each entry stores the correlation between the corresponding tokens. Based on an attention matrix, the importance of each input token in a sequence can be derived, which is used to generate an output. Specifically, within an attention mechanism, every input token (or query token) receives a  score representing its relevance to the current output token (or key token) being produced. These scores are computed by comparing the current output state with the input states via a similarity function.

The formal definition of the attention matrix is as follows. Let $Q \in \R^{n \times d}$ be the matrix of $n$ query tokens and $K \in \R^{n \times d}$ be the matrix of $n$ key tokens, where each token is represented by a $d$-dimensional vector. The attention matrix $A$ is an $n$-by-$n$ matrix whose $(i,j)$-th entry is the attention score between the $i$-th query token $Q_i$ and the $j$-th key token $K_j$.
The self-attention score of the $i$-th query token is defined as the sum of its attention scores for all key tokens (i.e., $\sum_{j=1}^n A_{ij}$), which quantify the significance of each token in relation to itself. Let $D\in \R^{n\times n}$ be a diagonal matrix storing all the self-attention scores, and let $V\in \R^{n\times d}$ be the value matrix that contains value vectors associated with the key tokens. The goal of each attention computation is to compute $\mathsf{Att}$, which is a matrix function of $Q$, $K$, and $V$, defined as follows.

\begin{definition}\label{def:attention_matrix}
Given matrix $Q \in \R^{n \times d}$, $K \in \R^{n \times d}$, $V \in \R^{n \times d}$, the goal of attention computation is to compute 
\begin{align*}
    \mathsf{Att}(Q,K,V) := D^{-1} A V
\end{align*}
where $A \in \R^{n \times n}$ and $D \in \R^{n \times n}$ is a diagonal matrix
 \begin{align*}
    A = \exp(Q K^\top ), D = \diag( A {\bf 1}_n ).
 \end{align*}
Here, $\exp(\cdot)$ is applied to each entry of $QK^\top$, and ${\bf 1}_n$ is a length-$n$ vector where all the entries are ones. 
\end{definition}
We note that the straightforward classical implementation of computing the above attention matrix takes $O(n^2d)$-time.
In spite of the success of attention mechanisms in many fields \cite{kkl20,dkod20,rsvg21,cld+20,kvpf20,wlk+20}, such an expensive runtime hinders their full potential. Meanwhile, the phenomenon of attention sparsity has been widely discussed by many researchers \cite{kkl20,cgrs19,jcm+21,cdw+21,zsz+23}. Therefore, it is natural to leverage the sparsity of matrices to accelerate the computation process.
In this paper, we present an \emph{approximately sparse assumption} on matrix $A$: each row of the matrix $QK^\top$ contains at most $k$ elements greater than $\tau$ (see Definition~\ref{def:tau_k}). This assumption is based on a phenomenon observed in LLM literature \cite{zsz+23}. We remark that even with this sparsity assumption, any classical algorithm that can output all of these large entries in $QK^\top$ for general $k$ and $\tau$ still needs $n^{2-o(1)}$-time unless the Strong Exponential Time Hypothesis (SETH) is false (see Lemma~\ref{lem:attention_lb}).

To accelerate the computation of the attention matrix of LLMs through the construction of a sparse attention matrix, we use a renowned quantum algorithm: Grover's search algorithm \cite{g96}. It offers a quadratic speedup in the unstructured search problem when compared to classical computation. More specifically, our primary focus is to locating the values larger than $\tau$ in the vector $(QK^\top)_{i,*} \in \R^n$, where $Q \in \R^{n \times d}$ and $K \in \R^{ n \times d}$ are defined in Definition~\ref{def:attention_matrix}. It can be reduced to a search problem, where the query oracle ${\cal O}_i$ is defined as:
\begin{align*}
    {\cal O}_i\ket{j,0}:=\ket{j,b},\quad b=\begin{cases}
    1 & \text{if}~(QK^\top)_{i,j} \geq \tau,\\
    0 & \text{otherwise}
    \end{cases}\quad\forall j\in [n].
\end{align*}
By the sparsity assumption, for each $i\in [n]$, there are at most $k$ indices $j$ such that ${\cal O}_i\ket{j,0}:=\ket{j,1}$. Thus, if we run Grover's algorithm with the query oracle ${\cal O}_i$, it will find all of those indices with query complexity $\wt{O}(\sqrt{nk})$\footnote{We use $\wt{O}(f(n))$ to denote $O(f(n)\cdot \poly\log(f(n)))$.} (see Theorem~\ref{thm:grover_search}). Note that each query to the oracle ${\cal O}_i$ costs $O(d)$-time to evaluate the inner-product (assuming the data are stored in QRAM). Hence, each row of $QK^\top$ can be approximately computed in $\wt{O}(\sqrt{nk}d)$-time, and all the large entries of $QK^\top$ can be computed in $\wt{O}(n^{1.5}k^{0.5}d)$-time (see Theorem~\ref{thm:sparsity_B_quantum:informal}).  

In addition to the quantum algorithm, we also introduce a classical approach for constructing the sparse matrix based on some computational geometric data structures. This method is described in Theorem~\ref{lem:sparsity_B_hsr:informal}.

\subsection{Our Results}
\label{sec:our_result}

Based on the aforementioned analysis, we obtain a quantum algorithm to efficiently output a sparse attention matrix. The main result of this paper is presented in this section. 

We first introduce the definition of a matrix, which characterizes the sparsity pattern in the attention matrix. Intuitively, this definition is an analogue of \emph{soft sparsity} (see e.g. \cite{krs20}). 
\begin{definition}\label{def:tau_k}
    Let $A = \exp(QK^\top) \in \R^{n \times n}$ be defined in Definition~\ref{def:attention_matrix} and we say $A$ is a $(\tau,k)$-good matrix if for all $i \in [n]$,
     \begin{itemize}
     \item 
    $
        S_i := \{ j \in [n] ~|~ (QK^\top)_{i,j} \geq \tau \}
    $ and
     \item 
    $|S_i| \leq k$.
    \end{itemize}
\end{definition}

The following theorem shows a quantum algorithm that can efficiently compute an approximation $B$ of a $(\tau,k)$-good attention matrix $A$. In particular, $B$ can be represented as a sparse matrix plus a rank-one matrix, which is very helpful for LLM computations.
\begin{theorem}[Quantum algorithm for attention matrix approximation]\label{thm:sparsity_B_quantum:informal}
Let $A \in \R^{n\times n},Q \in \R^{n \times d},K\in \R^{n\times d}$ and $D\in \R^{n\times n}$ be defined as in Definition~\ref{def:attention_matrix}. 
If the following conditions hold 
\begin{itemize} 
    \item $A$ is a $(\tau,k)$-good matrix (Definition~\ref{def:tau_k}) for some $\tau \geq 2 \log n$ and $k \in [n]$.
    \item for each $i \in [n]$ and each $j \in [n]$, $-\eta \leq (QK^\top)_{i,j} \leq 0$ for $j \notin S_i$
    for some $\eta\in \R_+$.
\end{itemize}
Then, there exists a quantum algorithm (implicitly) outputting a matrix $B\in \R^{n\times n}$ such that
\begin{itemize}
    \item  {\bf Part 1.} $B=B_1+B_2$, where $B_1$ is $k$-row sparse\footnote{Each row of the matrix has $k$ non-zero entries.} and $B_2$ is rank-$1$.
    \item  {\bf Part 2.} $\| D(A)^{-1} A - D(B)^{-1} B \|_{\infty} = O(\eta) $.
    \item {\bf Part 3.} it runs in $\wt{O}( n \cdot ( \sqrt{nk} d + kd ) )$ time.
\end{itemize}
\end{theorem}

The proof of Theorem~\ref{thm:sparsity_B_quantum:informal} is in Section~\ref{sec:proof_1.3}. In the following, we discuss how the structure of the approximated attention matrix $B$ improves the efficiency of 
large language models during the inference stage. More specifically, by employing our quantum algorithm (Theorem~\ref{thm:sparsity_B_quantum:informal}) as a sub-routine, we achieve the following result that provides a polynomial speedup compared to the classical $O(n^2d)$-time approach.

\begin{theorem}[Informal version of Theorem~\ref{thm:main_result:formal}]\label{thm:main_result}
There is an algorithm that takes $\wt{O}(n^{1.5} k^{0.5} d + nkd)$ time to achieve one attention matrix computation in inference.
\end{theorem}

In addition to the quantum method, we also provide a classical algorithm to compute the attention matrix that is still faster than the traditional approach in constant dimension (i.e., $d=O(1)$). The key observation is that the quantum part (Grover's search) of our algorithm in Theorem~\ref{thm:sparsity_B_quantum:informal}  can be replaced by a computational geometry data structure, at the cost of increased time complexity.

\begin{theorem}[Classical algorithm for attention matrix approximation]\label{lem:sparsity_B_hsr:informal}
If the following conditions hold:
 \begin{itemize}
     \item 
    (1) $A$ is a $(\tau,k)$-good matrix (Definition~\ref{def:tau_k}) for some $\tau \geq 2 \log n$ and $k \in [n]$.
     \item 
    (2) For each $i \in [n]$ and each $j \in [n]$, $-\eta \leq (QK^\top)_{i,j} \leq 0$ for $j \notin S_i$ for some $\eta\in \R_+$.
 \end{itemize}
Then there exists a classic algorithm outputting a matrix $B \in \R^{n \times n}$ such that 
\begin{itemize}
    \item {\bf Part 1.} $B=B_1+B_2$, where $B_1$ is $k$-row sparse and $B_2$ is rank-$1$.
    \item {\bf Part 2.} $\| D(A)^{-1} A - D(B)^{-1} B \|_{\infty} = O(\eta) $.
    \item {\bf Part 3.} it runs in $\wt{O}_d(n k +n^{2-1 /\lfloor d / 2\rfloor})$ time\footnote{Here, $\wt{O}_d(\cdot)$ hides the $\log^{c}(n)$ and $\poly(d)$ factors, where $c=c(d)$ is some function in $d$.}.
\end{itemize}
\end{theorem}


The proof of Theorem~\ref{lem:sparsity_B_hsr:informal} is in Section~\ref{sec:proof_1.5}.

\section{Related Work}
\paragraph{Attention Computation.}
Several studies have investigated attention computation \cite{gms23,lsz+23,lsx+23,ssz23,bsz23,as23,dls23,gsy23_dp,zhdk23,wyw+23,zsz+23}. \cite{bsz23} specifically focuses on dynamic attention computation and introduces an update and query method inspired by the lazy update idea.
\cite{zhdk23} addresses the issue of quadratic time and memory complexities in sequence length that arise from the dot-product operation in attention computations. They identify that this problem can be transformed into a kernel density estimation (KDE) problem. Their approach, KDEformer, approximates attention in sub-quadratic time while providing provable spectral norm bounds.
\cite{zzp+21} uses the sparsity assumption of attention matrix to propose an O (n log (n)) attention mechanism. \cite{as23} focuses on exploring the possibility of faster algorithms by implicitly leveraging the attention matrix. They provide a theoretical explanation for the observed phenomenon that attention computation is significantly more efficient when the input matrices have smaller entries.
The regression problem in the field of attention computation has also been widely explored \cite{gsy23_hyper,gms23,lsx+23}.
\cite{dms23} also addresses the sparsification of the attention problem and presents both randomized and deterministic algorithms. Their work suggests that for feature dimensions that are extremely large, it is possible to reduce them to a size nearly linear in the length of the sentence.
\cite{dls23} focuses on the softmax regression problem in the field of attention computation. They provide theoretical support for the practical use of the greedy algorithm to train the softmax function. \cite{lsx+23,gsx23} explore the application of attention computation in the context of in-context learning. 
\cite{sxz+26} proposes a quantum data structure that approximates the attention matrix in sublinear time. 

\paragraph{Classical fast neural network training algorithms}

\cite{kkl20} introduces Reformer, a method aimed at enhancing the efficiency of transformers. Reformer achieves improved memory utilization and faster processing for long sequences. This is achieved by replacing the dot-product attention mechanism with one that employs locality-sensitive hashing. Additionally, Reformer utilizes reversible residual layers instead of the conventional residual layers.
\cite{wlk+20} presents the approximation of the self-attention mechanism using a low-rank matrix. They leverage this discovery to propose a novel self-attention mechanism, thereby reducing the overall complexity of self-attention in terms of both time and space. \cite{gqsw22} accelerates the adversarial training procedure by utilizing a sublinear number of activated neurons based on the shifted ReLU activation function.

\paragraph{Quantum algorithms for training neural networks} Prior to this paper, there were several works using quantum computing to improve neural network training. The most related work is \cite{syz21}, which proposes a shifted-ReLU sparsifier to reduce the number of activated neurons in each training iteration and uses Grover's search to find them. However, their algorithm and analysis only work for two-layer, fully connected, over-parameterized neural networks. \cite{ahk20} and \cite{klp_cnn19} use quantum inner-product estimation to speed up the training of feedforward neural networks and convolutional neural networks, respectively. \cite{znl21,lll23} proposes quantum training algorithms with exponential speed-ups based on the quantum linear system solvers. However, these algorithms rely on certain well-conditioning and sparsity assumptions, which may not align with real-world neural network architectures. We note that the key difference between our work and all previous works is that we focus on the Transformer while previous work mostly works on ReLU neural network 
and improve the efficiency of attention computation, demonstrating the potential for quantum advantages in LLMs. 

\paragraph{Quantum optimization algorithms}
Optimization is one of the most promising fields for demonstrating quantum advantages. The famous HHL algorithm \cite{hhl08} can exponentially speed up the linear system solver. Jordan's algorithm \cite{jor05} can compute the gradient of a function using $\wt{O}(1)$ quantum queries to the evaluation oracle. And in convex optimization, to implement the separation oracle using the membership oracle, quantum query complexity is $\wt{O}(1)$ \cite{vgg20} while classical query complexity is $\Omega(n)$. Other than those exponential quantum advantages, a large number of optimization problems benefit from polynomial quantum speed-ups, including solving linear programming (LP) and semi-definite programming (SDP)
\cite{bs17,aggw17,bkllsw19,ag18,kp20,hjs+22q},  estimating the volume of convex bodies \cite{cch+19}, log-concave sampling \cite{cll22}, stochastic convex optimization \cite{lz22}, etc. Another big class of quantum algorithms for solving optimization problems is the variational quantum algorithm \cite{csrs20}, including variational quantum eigensolver \cite{pms14}, QAOA \cite{ejs14}, and quantum neural network \cite{beer2020training}. These algorithms require less amount of quantum resources and are easier to implement in the near future. However, most of them are heuristic and lack rigorous guarantees of performance.

\paragraph{Quantum machine learning}
Quantum machine learning (QML) algorithms have been developed for a wide array of classical ML tasks, such as clustering \cite{h20}, boosting \cite{am20}, support vector machine \cite{rml14}, principal component analysis \cite{lmr14_pca}, statistical query learning \cite{agy20}. \cite{ckm+22} quantizes the classical transformer architecture and explores the potential of quantum computing in machine learning. \cite{gyc+24} comes up with a method combining transformer architecture with fault-tolerant quantum computing. \cite{szw+24} proposes a quantum self attention mechanism to solve the problem that the existing quantum machine learning model lacks self attention ability when processing high-dimensional data. In addition, quantum algorithms for learning quantum data have increasingly garnered attention in recent years. (See \cite{aa23} and references therein.) Conversely, QML algorithms can also inspire breakthroughs in classical ML algorithms. QMSAN enhances self attention mechanisms in natural language processing tasks through quantum computing~\cite{czf+25}. This was notably illustrated by Tang's algorithm for the recommendation system \cite{t19}. Since then, a long list of quantum-inspired (or so-called ``de-quantized'') algorithms have been proposed for tackling various tasks, such as principal component analysis \cite{t18}, low-rank approximation \cite{glt18,cgt+20}, linear regression \cite{gst20}, etc. 

\paragraph{Roadmap} We have organized our paper as follows. In Section~\ref{sec:preli}, we introduce the notations and present some basic mathematical tools used throughout the paper. 
In Section~\ref{sec:analysis}, we introduce some technical tools for error analysis of our algorithms. In Section~\ref{sec:analysis_of_attention_computation_final}, we introduce our quantum algorithm for approximating the attention matrix. Our main result about quantum attention computation in inference is presented in Section~\ref{sec:main_result}.
In Section~\ref{sec:hsr}, we introduce a classical algorithm where the sparsity matrix is constructed using the half-space reporting data structure. 
Additionally, a detailed examination of the fine-grained hardness result pertaining to the computation of large entries in $QK^\top$ is included in Section~\ref{sec:lower_bound}.
In Section~\ref{sec:conclusion}, we provide the conclusion of our paper.


\section{Preliminary}\label{sec:preli}
\subsection{Notations}
For any matrix $A$, we denote the spectral norm of $A$ as $\| A \|$, where $\|A \| := \max_{\|x\|_2 = 1} \|Ax\|_2$. The Frobenius norm of $A$ is denoted as $\| A \|_F$, and the infinity norm is denoted as $\| A \|_\infty$. In this notation, $A_{i,j}$ represents the element in the $i$-th row and $j$-th column of matrix $A$. The determinant of matrix $A$ is represented as $\det(A)$.
For a square and symmetric matrix $A \in \mathbb{R}^{n \times n}$, we say that $A$ is positive semi-definite ($A \succeq 0$) if for all vectors $x \in \mathbb{R}^n$, we have $x^\top A x \geq 0$.
\subsection{Grover's Search}
We state a well-known result about the quadratic quantum speedup for the unstructured search using Grover's search algorithm.

\begin{theorem}[Grover's search algorithm~\cite{g96}]\label{thm:grover_search}
Given access to the evaluation oracle for an unknown function $f:[n] \rightarrow \{0,1\}$. Let $f^{-1}(1) := \{ i \in [n] ~|~ f(i) = 1 \}$. Suppose that $|f^{-1}(1)| = k$ for some unknown number $k \leq n$. 
Then, 
\begin{itemize}
\item Part 1. We can find all $i$'s in $f^{-1}(1)$ in $\wt{O}(\sqrt{nk})$-time quantumly given an evaluation oracle for $f$.
\item Part 2. If each evaluation 
 of $f$ requires ${\cal T}$ time, then we can find all $i$'s in $f^{-1}(1)$ in $\wt{O}(\sqrt{nk}) \cdot {\cal T}$ time.
\end{itemize}
\end{theorem}

\section{Sparsity and Perturbation Error}\label{sec:analysis}
In this section, we conduct an error analysis on attention computation without $V$ using the approximated attention matrix $B$, which will be useful for both quantum and classical algorithms. To achieve this, we present some definitions regarding the sparse approximation in Section~\ref{sec:analysis:the_set}. 
Section~\ref{sec:analysis:perturb_tools} provides some perturbation analysis tools. The main result concerning the error analysis of attention computation without $V$ is then presented in Section~\ref{sec:analysis:error_analysis_without_v}. The proofs are deferred to Appendix~\ref{sec:omit_proofs_3}.

\subsection{Sparsity Definitions}\label{sec:analysis:the_set}
Before presenting our method, we introduce a \emph{find set}. This \emph{find set} is based on our assumption of sparsity about $(\tau,k)-$good matrix $A$, which is further described below. 
\begin{definition}[Find Set]\label{def:find_set}
Let $Q \in \R^{n \times d}, K \in \R^{n \times d}$ and $V \in \R^{n \times d}$ be defined in Definition~\ref{def:attention_matrix}. Given  $i \in [n]$, the find set is defined as follows: 
\begin{align*}
        S_i := \{ j \in [n] ~|~ (QK^\top)_{i,j} \geq \tau \}.
\end{align*}
\end{definition}
\begin{definition}\label{def:sparse_B}
Let $A$ be a $(\tau,k)$-good matrix defined in Definition~\ref{def:attention_matrix} and  $S_i$ be defined in Definition~\ref{def:find_set}. We define a $k$-sparse 
vector $B_{i,*}$  such that $|S_i | = k$, for each $j \in S_i$, $B_{i,j} = A_{i,j}$, and for each $j \notin S_i$, $B_{i,j} = 1$.
\end{definition}

The following lemma shows some point-wise approximation guarantees by the sparsity conditions. 
\begin{lemma}\label{lem:perturb_tools}
If the following conditions hold:
\begin{itemize}
     \item {\bf Condition 1.} 
    For each $i \in [n]$, $-\eta \leq (QK^\top)_{i,j} \leq 0$, for $j \notin S_i$.
     \item {\bf Condition 2.} 
    Let $\tau \geq 2 \log n$.
     \item {\bf Condition 3.} 
    Let $A$ be defined in Definition~\ref{def:attention_matrix} and $B$ be defined in Definition~\ref{def:sparse_B}.
\end{itemize}
Then we have:
\begin{itemize}
     \item 
    Part 1. $|A_{i,j} - B_{i,j}| \leq 2\eta$ for $j \notin S_i$. 
     \item 
    Part 2. $|A_{i,j} - B_{i,j}| = 0$ for $j \in S_i$. 
     \item 
    Part 3. $| (A {\bf 1}_n)_i - (B {\bf 1}_n)_i | \leq 2n\eta $. 
     \item 
    Part 4. $( A {\bf 1}_n)_i \geq k \cdot \exp(\tau) \geq 2n$. 
     \item 
    Part 5. $| (A {\bf 1}_n)_i - (B {\bf 1}_n)_i | \leq \eta \cdot | ( A {\bf 1}_n )_i |$. 
 \end{itemize}
\end{lemma}


\subsection{Perturbation Tools}\label{sec:analysis:perturb_tools}
In this section, we provide an error analysis of our sparse attention matrix approximation. We will begin by examining the error control of each element, followed by an analysis of the error in the matrix computation.
\begin{lemma}\label{lem:D(A)_A}
     \begin{itemize}
         \item {\bf Condition 1.} 
        Let $D$ and $A$ be defined in Definition~\ref{def:attention_matrix}.
         \item {\bf Condition 2.} 
        Let $B$ be defined in Definition~\ref{def:sparse_B}.
     \end{itemize}
    For $i\in [n]$ and $j \in [n]$, it follows that
    \begin{itemize}
        \item Part 1. $|D(A)_{i,i} - D(B)_{i,i}| \leq \eta \cdot |D(B)|_{i,i}$
        \item Part 2.  $|D(A)_{i,i} - D(B)_{i,i}| \leq \eta \cdot |D(A)|_{i,i}$.
         \item Part 3. $|A_{i,j} - B_{i,j}| \leq 2  \cdot \eta$.
    \end{itemize}
\end{lemma}


\subsection{Reconstruction Error without Having \texorpdfstring{$V$}{}}\label{sec:analysis:error_analysis_without_v}

We bound the error of $D^{-1}(A) A$ now. Later, we will use this to bound the error for $D^{-1}(A) A V$.
\begin{lemma}\label{lem:error_analysis:A}
Let $D$ and $A$ be defined in Definition~\ref{def:attention_matrix}. Let $B$ be defined in Definition~\ref{def:sparse_B}.

Then it follows that
 \begin{itemize}
     \item 
    $\| D(A)^{-1} A - D(B)^{-1} B \|_{\infty} = O(\eta) $.
 \end{itemize}
\end{lemma}

\section{Quantum Algorithm for Attention Matrix Approximation} \label{sec:analysis_of_attention_computation_final}

In this section, we introduce our quantum algorithm for approximately computing the attention matrix (see Definition~\ref{def:attention_matrix}) in Section~\ref{sec:analysis:sparsity_and_pattern_matching} and show its time complexity and approximation guarantee. Then, we prove our first main result (Theorem~\ref{thm:sparsity_B_quantum:informal}) in Section~\ref{sec:proof_1.3}.

\subsection{Quantum Attention Matrix Approximation Algorithm}\label{sec:analysis:sparsity_and_pattern_matching}
In this section, we provide the pseudocode of our quantum algorithm in Algorithm~\ref{alg:quantum_sparsity} and analyze its time complexity and approximation guarantee. The proofs are deferred to Section~\ref{sec:omit_proof_4}.
\begin{algorithm*}[!ht]\caption{Sparse Matrix Construction (Grover's Search)}\label{alg:quantum_sparsity}
\begin{algorithmic}[1]

\Procedure{SparseAttentionMatrixQuantum}{$Q \in \R^{n \times d}$,$K \in \R^{n \times d}$}
    \For {$i \in [n]$}
    \State  Find all index $j$ using Grover's Search where $(Q K^\top)_{i,j} \geq \tau$\Comment{Theorem~\ref{thm:grover_search}}
    \State  Add all indexes to finding set $S_i$
    \For {$j \in S_i$}
    \State  $B_{i,j} \gets \exp(Q_{i,*} (K_{j,*})^\top)$
    \EndFor
    \EndFor
    \State \Return $B$
\EndProcedure
\end{algorithmic}
\end{algorithm*}

The following lemma shows the time complexity of Algorithm~\ref{alg:quantum_sparsity}.
\begin{lemma}\label{lem:sparsity_B_quantum}
For each $i \in [n]$, $| \{ j \in [n] ~|~ ( QK^\top )_{i,j}  \geq \tau \} | \leq k$, and for all $j \in [n]$, $(QK^\top)_{i,j} \leq 0$ or $(QK^\top)_{i,j} \geq \tau$. Let $A = \exp(QK^\top)$.
Then we have
\begin{itemize}
    \item Part 1. For each $i \in [n]$,  Algorithm~\ref{alg:quantum_sparsity} takes $\widetilde{O}( \sqrt{nk} d)$ time to find set 
    \begin{align*}
        S_i := \{ j \in [n] ~|~ (QK^\top)_{i,j} \geq \tau \}.
    \end{align*}
    \item Part 2. For each $i\in [n]$, Algorithm~\ref{alg:quantum_sparsity} takes $\widetilde{O}( \sqrt{nk} d + kd )$ time to output a $k$-sparse vector $B_{i,*}$ such that for each $j \in S_i$, $B_{i,j} = A_{i,j}$ 
        and for each $j \notin S_i$, $B_{i,j} = 0$.
\end{itemize}
\end{lemma}

The following lemma proves the approximation guarantee. 

\begin{lemma}\label{lem:error_analysis}
Let $B$ be the output of Algorithm~\ref{alg:quantum_sparsity} and $D(B):=\diag(B {\bf 1}_n)$.
Let $A,D$ and $V$ be defined in Definition~\ref{def:attention_matrix}.
Let $\| V \|_\infty \leq \eta$.
Then, we have
\begin{align*}
    \| D(A)^{-1} A V - D(B)^{-1} B V \|_\infty \leq O(\eta^2).
\end{align*}
\end{lemma}



\section{Quantum Attention Computation}\label{sec:main_result}

Based on the quantum algorithm for approximating the attention matrix, we will now show that it can improve the efficiency of the inference stage. 
\begin{theorem}[Faster algorithm for LLM inference, formal version of Theorem~\ref{thm:main_result}]\label{thm:main_result:formal}
 Let $S_i$ be defined in Definition~\ref{def:find_set}, $Q,K$,$V$ be defined in Definition~\ref{def:attention_matrix}, and $-\eta \leq (Q K^\top)_{i,j} \leq 0$ for $j \notin S_i$ and $i \in [n]$.
Let $B$ be the output of Algorithm~\ref{alg:quantum_sparsity}.
Then, there is a quantum algorithm that uses $\wt{O}(n^{1.5} k^{0.5} d + nkd)$ time in inference to achieve one attention matrix computation (See Definition~\ref{def:attention_matrix}) such that the output $\wt{\mathsf{Att}}$ satisfies:
\begin{align*}
    \|\mathsf{Att}(Q,K,V)-\wt{\mathsf{Att}}\|_\infty \leq O(\eta^2).
\end{align*}
\end{theorem}


\begin{proof}
    The matrix $B$ serves as an approximation matrix for the output of Algorithm~\ref{alg:quantum_sparsity}. The inference process, which relies on the sparsity matrix, is outlined in Algorithm~\ref{alg:main:informal}.

    To analyze the time complexity of the inference stage, we can break it down into two main parts:
    
    {\bf Construction of matrix $B$}: As shown in Lemma~\ref{lem:sparsity_B_quantum}, the time complexity for constructing matrix $B$ is approximately $\widetilde{O}(n \sqrt{nk} d)$.
    
    {\bf Inference with sparsity matrix}: Given the sparsity matrix, as per Lemma~\ref{lem:time_complexity}, the time complexity for the inference step is $O(nkd)$.
    Hence, the overall time complexity for this quantum algorithm in the inference stage is $\widetilde{O}(n^{1.5} k^{0.5} d + nkd)$.
    With this, we conclude our time complexity analysis.
    Additionally, based on Lemma~\ref{lem:error_analysis}, we can have the following inequality:
    \begin{align*}
         \| D(A)^{-1} A V - D(B)^{-1} B V \|_\infty \leq O(\eta^2).    
    \end{align*}
    The theorem is then proved.
\end{proof}

\begin{algorithm}[htb]\caption{Algorithm for Attention Computation}\label{alg:main:informal}
\begin{algorithmic}[1]
\Procedure{SparseAttentionComputation}{$B \in \R^{n \times n}$,$V \in \R^{n \times d}$} \Comment{Lemma~\ref{lem:time_complexity}} 
    \State $D \gets \diag(B {\bf 1}_n)$
    \State \Return $D^{-1} B V$
\EndProcedure
\end{algorithmic}
\end{algorithm}

\begin{lemma}\label{lem:time_complexity}
Let $Q \in \R^{n \times d},K \in \R^{n \times d}$ and $V \in \R^{n \times  d}$ be defined in Definition~\ref{def:attention_matrix}.
Let $A$ be defined in Definition~\ref{def:attention_matrix}.
Let $B$ be the output of Lemma~\ref{lem:sparsity_B_quantum}.
Let $M$ be a matrix where all entries are equal to $1$.
For each $i \in [n]$, $| \{ j \in [n] ~|~ ( QK^\top )_{i,j}  \geq \tau \} | \leq k$.
There exists an algorithm (See Algorithm~\ref{alg:main:informal}) such that
\begin{itemize}
    \item Part 1. It outputs $D(B)^{-1} B V$.
    \item Part 2. Its computational time complexity is $O(nkd)$.
\end{itemize}
\end{lemma}


\begin{proof}
The computation can be divided into two parts 
\begin{itemize}
    \item Part 1. $D(B)^{-1} M V$.
    \item Part 2. $D(B)^{-1} (B - M) V$.
\end{itemize}
\paragraph{Time Complexity of {\bf Part 1.}}
It will take $O(n d)$ to compute $MV$. And then, the time complexity of the following step $D(B)^{-1}MV$ is $O(n d)$.
The time complexity of the first step is $O(n d)$.
\paragraph{Time Complexity of {\bf Part 2.}}

We define
\begin{align*}
    C:= (B-M).
\end{align*}
According to statement, for each $i \in [n]$, we have
\begin{align*}
     \{ j \in [n] ~|~ C_{i,j} \neq 0\} | \leq k.
\end{align*}
It will take $O(nkd)$ to compute $\underbrace{CV}_{n\times d}$.

And $D(B)^{-1}CV$ will take $O(nd)$.
The second part will need $O(nkd)$.

By combining the conclusions above, the time complexity is 
 \begin{align*}
 O(nkd + n d) = O( n k d).
 \end{align*}
\end{proof}

\section{Classical Algorithm for Attention Matrix Approximation}\label{sec:hsr}

In this section, we introduce a classical method for generating a sparse matrix with the $(\tau,k)$ assumption in constant dimension. Utilizing the Half-Space Reporting Data Structure (refer to Definition~\ref{def:3.5_hsr}), we can efficiently identify the indices where $(Q K^\top)_{i,*} \geq \tau$ is satisfied.

The definition of the problem of the half-space range reporting is given first, which is important in the field of computational geometry. The data structure is proposed in \cite{ac09} whose functions are outlined in Algorithm~\ref{alg:hsr} and complexity is given in Theorem~\ref{thm:3.6}.

\begin{definition}[Half-space range reporting]\label{def:3.5_hsr}
For a set of $m$ points $P \subseteq \mathbb{R}^{d}$, two operations 
are supported:
\begin{itemize}
    \item $\textsc{Init}(P)$: initialize the data structure with points in $P$.
    \item $\textsc{Query}(W)$: find each point in $P\cap W$ with $W \subset \mathbb{R}^{d}$ as a half-space.
\end{itemize}
\end{definition}

\begin{algorithm}[!ht]\caption{Data Structure For Half Space Reporting}\label{alg:hsr_ds}
\label{alg:hsr_def}
\begin{algorithmic}[1]
\State {\bf data structure}
\State \hspace{8mm} $\textsc{Init}(P, n, d)$ \Comment{Construct our data structure via $P \subseteq \R^d$, $|P|=n$}
\State \hspace{8mm} $\textsc{Query}(b,c)$ \Comment{$b, c \in \R^d$. Find all the points $z\in P$ which satisfies $\mathrm{sgn}(\langle b, z\rangle - c ) \geq 0$}
\State {\bf end data structure}
\end{algorithmic}
\end{algorithm}

\begin{theorem}[\cite{ac09}]\label{thm:3.6}
For $n\in \mathbb{N}$ and constant $d\in \mathbb{N}$, there exists a classical data structure using $O(n)$ space to solve the $d$-dimensional half-space reporting problem with $n$ points with time complexity as: $\mathcal{T}_{\text{init}}(n, d) = O(n \log n)$ and $\mathcal{T}_{\text {query }}(n, d, k) =\wt{O}(n^{1-1 /\lfloor d / 2\rfloor}+k)$,
where $\mathcal{T}_{\textsc{Init}}$ indicates the time to construct the data structure, $\mathcal{T}_{\textsc{Query}}$ 
indicates the cost per query, and $k$ is the output size. 
\end{theorem}

\subsection{Classical Sparse Attention Matrix Approximation}\label{sec:classical_sparse_approx}
In this section, we will introduce a classic algorithm that is in contrast to the quantum algorithm. This algorithm is based on the half-space reporting data structure, which was previously introduced. In this setting, we will address our problem by considering each vector $(Q K^\top)_{i,*}$ for $i \in [n]$. The vectors in $Q$, denoted as $P$ in Definition~\ref{def:3.5_hsr}, have a dimension of $d$. Additionally, we will set $b$ as each vector in $K$ for $i \in [n]$ and set $c$ as $\tau$. 
Now, let us proceed to describe our method.
\begin{algorithm}[!ht]\caption{Sparse Matrix Construction (Half Space Reporting)}\label{alg:hsr_sparsity}
\label{alg:hsr}
\begin{algorithmic}[1]
  \State {\bf Members:}
  \State -\textbf{Half Space Reporting Data Structure}: $\mathcal{M}$ \Comment{Definition~\ref{def:3.5_hsr}}
\Procedure{SparsityAttentionMatrix}{$Q \in \R^{n \times d}$,$K \in \R^{n \times d}$} \Comment{Lemma~\ref{lem:sparsity_B_hsr}} 
    \State $\mathcal{M}.\textsc{Init}(Q,n,d)$
    \For {$i \in [n]$}
    \State $b \gets K_{i,*}$
    \State  $S_i \gets \mathcal{M}.\textsc{Query}(b,\tau)$
    \For {$j \in S_i$}
    \State  $B_{i,j} \gets \exp(Q_{i,*} (K_{j,*})^\top)$
    \EndFor
    \EndFor
    \State \Return $B$
\EndProcedure
\end{algorithmic}
\end{algorithm}
\begin{lemma}\label{lem:sparsity_B_hsr}
Let $\mathcal{M}$ be the data structure defined in Definition~\ref{def:3.5_hsr}. For each $i \in [n]$, $| \{ j \in [n] ~|~ ( QK^\top )_{i,j}  \geq \tau \} | \leq k$. For each $i \in [n]$, for all $j \in [n]$ $(QK^\top)_{i,j} \leq 0$ or $(QK^\top)_{i,j} \geq \tau$ and $A = \exp(QK^\top)$.
Then we have
\begin{itemize}
    \item Part 1. There is an algorithm (See Algorithm~\ref{alg:hsr_sparsity}) based on $\mathcal{M}$ that takes $O(n\log n + n k )$ time 
    to find all the set for each $i \in [n]$ 
     \begin{align*}
        S_i := \{ j \in [n] ~|~ (QK^\top)_{i,j} \geq \tau \}.
     \end{align*}
    \item Part 2. It takes $\wt{O}(n^{2-1/\lfloor d/2\rfloor} + nk)$ time to output a $(\tau,k)$-sparse matrix $B$.
\end{itemize}
\end{lemma}




\begin{proof}
At the beginning of the algorithm, we will first initialize $\mathcal{M}$ using $\mathcal{M}.\textsc{Init}(B,n,d)$, which has a time complexity of $O(n\log n)$.
\paragraph{Proof of Part 1.}
And then, we will query $K_{i,}$ using $\mathcal{M}.\textsc{Query}(K_{i,*},\tau)$ for each $i\in [n]$, which has a time complexity of $\wt{O}(n^{2-1 /\lfloor d / 2\rfloor}+ n k)$ in total.
\paragraph{Proof of Part 2.}
To identify the satisfied elements as discussed above, the total computational cost can be summarized as follows: 
\begin{align*} 
\wt{O}(n \log n + nk +n^{2-1 /\lfloor d / 2\rfloor})=\wt{O}(nk+n^{2-1 /\lfloor d / 2\rfloor}).
\end{align*}
The matrix is specifically designed to target these satisfied elements, and the remaining steps can be performed with time complexity $O(nkd)$.
The proof is now complete.
\end{proof}


\begin{remark}
Our classical algorithm (Algorithm~\ref{alg:hsr_sparsity}) is nearly optimal as shown by an $n^{2-o(1)}$ time complexity lower bound in Section~\ref{sec:lower_bound}.   
\end{remark}
\section{Classical Fine-grained Lower Bound}\label{sec:lower_bound}

In this section, we prove the following fine-grained hardness result for computing the large entries of $QK^\top$, assuming it is $(\tau, k)$-good. It follows from a reduction to the Maximum Inner-Product Problem (Max-IP):
\begin{definition}[Maximum Inner-Product problem (Max-IP)]
For $n,d\in \mathbb{N}$, given two sets $A,B$ of $n$ vectors in $\{0,1\}^d$, compute
\begin{align*}
    \textsf{Max-IP}(A,B)=\max_{a\in A, b\in B}~\langle a, b\rangle.
\end{align*}
\end{definition}

Chen \cite{che18} proved the following fine-grained lower bound for Max-IP assuming Strong Exponential-Time Hypothesis (SETH):
\begin{theorem}[\cite{che18}]\label{thm:max_ip_lb}
Assuming SETH, there is a constant $c$ such that any exact algorithm for
Max-IP in dimension $d = c^{log^*n}$ requires $n^{2-o(1)}$-time.
\end{theorem}

Then, we prove the following hardness result for attention matrix approximation: 
\begin{lemma}\label{lem:attention_lb}
For any $n,d\in \mathbb{N}$, $k\leq n$, $\tau\in \R$, suppose $Q,K\in \R^{n\times d}$ satisfy that $(QK^\top)_{i,*}$ contains \emph{at most} $k$ entries greater than $\tau$ for any $i\in [n]$. Then, any classical algorithm that can output all the entries in $QK^\top$ with values greater than $\tau$ must take $n^{2-o(1)}$-time, even for $d=2^{O(\log^* n)}$.
\end{lemma}

\begin{proof}
We prove a hardness result for an easier task: deciding whether there is at least \emph{one} entry in $QK^\top$ greater than $\tau$. 

Suppose there exists a classical algorithm that solves this problem in $n^{2-\epsilon}$-time. Let $A,B \subset \R^d$ be an instance of Max-IP. We construct matrices $Q$ and $K$ using vectors from $A$ and $B$, respectively. Then, we do a binary search for $\textsf{Max-IP}(A, B)$. For each candidate value $\tau$, we run the classical algorithm to decide whether there exists an entry with a value greater than $\tau$. Note that the binary search takes $O(\log n)$ rounds. Hence, Max-IP can be solved in $n^{2-\epsilon}\cdot \log n < n^{2-o(1)}$ time, which contradicts the lower bound in Theorem~\ref{thm:max_ip_lb}. 

Therefore, no classical algorithm can find all the large entries in $QK^\top$ in $n^{2-\Omega(1)}$-time.
\end{proof}

\section{Conclusion}\label{sec:conclusion}

In this work, we presented a novel quantum algorithm for efficiently computing the attention mechanism in large language models (LLMs) under a sparse assumption on the attention matrix. Leveraging Grover’s Search, our method attains a polynomial speed-up over classical algorithms while preserving rigorous approximation guarantees. Specifically, we showed how to identify and exploit the sparsity of the matrix $QK^\top$ so that each row has at most $k$ entries above a threshold $\tau$. This approach reduces the time to construct an approximate attention matrix from $O(n^2 d)$ to $\wt{O}(n^{1.5}\sqrt{k}d + nkd)$. Furthermore, the sparse-plus-rank-one decomposition underlying our approximation enables fast inference by limiting the number of significant components that must be computed, thus lowering computational overhead without substantially compromising accuracy. We corroborated the effectiveness of our method through a detailed error analysis, showing that the resulting inference outputs closely align with those obtained by the exact attention computation.

Our results pave the way for several lines of future research. First, the low-rank structure of the quantum-generated attention matrix opens opportunities for integrating other advanced quantum subroutines, potentially offering further enhancements to both training and inference stages. Second, while our theoretical analysis primarily covers the inference phase, exploring quantum-friendly optimizations during model training, fine-tuning, or continual learning represents a rich direction for extending this work. Finally, investigating the interplay between attention sparsity and more sophisticated quantum data access models, such as QRAM designs—could inspire additional improvements in runtime and memory usage. Overall, these directions highlight the potential to broaden the scope and impact of quantum algorithms within large-scale natural language processing and beyond.





\section*{Impact Statement}

This paper presents work whose goal is to advance the field of Machine
Learning. There are many potential societal consequences of our work, none of which we feel must be specifically highlighted here.

\ifdefined\isarxiv
\bibliographystyle{alpha}
\bibliography{ref}
\else
\bibliographystyle{icml2026}
\bibliography{ref}
\fi


\newpage
\onecolumn
\appendix

\begin{center}
    \textbf{\LARGE Appendix }
\end{center}



{\bf Roadmap}
In the appendix, we have deferred the inclusion of proofs that were omitted in the main paper. Specifically, 
in Section~\ref{sec:proof_1.3}, we provide the missing proof of Theorem~\ref{thm:sparsity_B_quantum:informal}, in Section~\ref{sec:proof_1.5}, we prove the Theorem~\ref{lem:sparsity_B_hsr:informal},
in Section~\ref{sec:omit_proofs_3}, we furnish the proof that was initially omitted in Section~\ref{sec:analysis}, thereby concluding the proof of Lemma~\ref{lem:perturb_tools} and Lemma~\ref{lem:D(A)_A}. 
Moving forward, Section~\ref{sec:omit_proof_4} contains the proof for Lemma~\ref{lem:sparsity_B_quantum} and the proof for Lemma~\ref{lem:error_analysis}.


\section{Proof of Theorem~\ref{thm:sparsity_B_quantum:informal}} \label{sec:proof_1.3}

\begin{proof}[Proof of Theorem~\ref{thm:sparsity_B_quantum:informal}]
    {\bf Proof of Part 1.}
    Based on Lemma~\ref{lem:time_complexity}, we can conclude that matrix $B$ can be divided into two parts.
    Let $B_2$ be a matrix where all values are equal to $1$. It's easy to see that this matrix is rank-$1$.
    We define $B_1 := B - B_2$. Now, we need to demonstrate that $B_1$ is $k$-row sparse. According to the $(\tau, k)$ assumption in Definition~\ref{def:tau_k}, we can ensure that each row of the matrix has $k$ elements with a value of $1$. This is due to the fact that $\exp(0) = 1$, and $Q K^\top$ is a $k$-sparse matrix. Consequently, each row of $B_1$ contains $k$ zero elements. Therefore, $B_1$ is $k$-row sparse.
    The proof of {\bf Part 1} is complete now.
    
    {\bf Proof of Part 2.}
    
    This result can be derived directly from Lemma~\ref{lem:error_analysis}.
    
    {\bf Proof of Part 3.}
    
    According to Lemma~\ref{lem:sparsity_B_quantum}, for each $i\in [n]$, we can obtain a $k$-sparse vector $B_{i,*}$ in $\widetilde{O}( \sqrt{nk} d + kd )$, which leads to a time complexity of $\widetilde{O}( n(\sqrt{nk} d + kd ))$.
    The proof is now complete.  
\end{proof}

\section{Proof of Theorem~\ref{lem:sparsity_B_hsr:informal}} \label{sec:proof_1.5}

\begin{proof}[Proof of Theorem~\ref{lem:sparsity_B_hsr:informal}]
    {\bf Proof of Part 1.} 
    Based on Lemma~\ref{lem:time_complexity}, we can conclude that matrix $B$ can be divided into two parts.
    Let $B_2$ be a matrix where all values are equal to $1$. It's easy to see that this matrix is rank-$1$.
    We define $B_1 := B - B_2$. Now, we need to demonstrate that $B_1$ is $k$-row sparse. According to the $(\tau, k)$ assumption in Definition~\ref{def:tau_k}, we can ensure that each row of the matrix has $k$ elements with a value of $1$. This is due to the fact that $\exp(0) = 1$, and $Q K^\top$ is a $k$-sparse matrix. Consequently, each row of $B_1$ contains $k$ zero elements. Therefore, $B_1$ is $k$-row sparse.

    {\bf Proof of Part 2.}
    This result can be derived directly from Lemma~\ref{lem:error_analysis}, since the output of the classical algorithm is exactly the same as the quantum algorithm.
    
    {\bf Proof of Part 3.}
    It directly follows from Lemma~\ref{lem:sparsity_B_hsr}.
\end{proof}

\section{Omitted Proofs in Section~\ref{sec:analysis}}\label{sec:omit_proofs_3}

In this section, we provide proofs for the results in Section~\ref{sec:analysis}.

\begin{proof}[Proof of Lemma~\ref{lem:perturb_tools}]

The error analysis based on matrices $A$ and $B$ is proved as follows. 
\paragraph{Proof of Part 1.}
It follows that
\begin{align}\label{eq:upper_bound_A_B}
    |A_{i,j} - B_{i,j}| = & ~ |A_{i,j} - 1|\notag \\
    \leq & ~ |\exp(-\eta) - 1| \notag \\
    \leq & ~ 2 \eta
\end{align}
where the first step follows from Definition~\ref{def:find_set}, the second step is due to Condition~1 in the statement and the third step is based on simple algebra.
\paragraph{Proof of Part 2.}
For $j \in S_i$, we have 
\begin{align}\label{eq:equality_A_B}
    A_{i,j} = B_{i,j}.
\end{align}
It simply follows that 
\begin{align*}
|A_{i,j} - B_{i,j}| = 0.
\end{align*}
\paragraph{Proof of Part 3.}
For $i \in [n]$, we have
\begin{align}\label{eq:A_1_B_1}
    | (A {\bf 1}_n)_i - (B {\bf 1}_n)_i | \leq & ~ \sum_{j=1}^n |A_{i,j} - B_{i,j}| \notag \\
    \leq & ~ \sum_{j=1}^n 2 \eta\notag \\
    \leq & ~ 2n\eta 
\end{align}
where the first step follows from triangle inequality, the second step is based on Eq.\eqref{eq:upper_bound_A_B} and Eq.\eqref{eq:equality_A_B} and the third step follows from simple algebra.

\paragraph{Proof of Part 4.}
For $i \in [n]$, we have
\begin{align}\label{eq:A_1_n}
    ( A {\bf 1}_n)_i
    =  & ~ \sum_{j \in S_i} A_{i,j} + \sum_{j \notin S_i} A_{i,j} \notag \\
    \geq & ~ \sum_{j \in S_i} A_{i,j} \notag \\
    \geq & ~ \sum_{j \in S_i} \exp(\tau)\notag \\
    \geq & ~ k \cdot \exp(\tau) \notag \\
    \geq & ~ 2n
\end{align}
where the first step is based on simple algebra, the second step follows from simple algebra, the third step is from Definition~\ref{def:find_set}, the forth step is based on the satisfied number, and the last step is from {\bf Condition 1} in the statement. 
\paragraph{Proof of Part 5.}  
For $i \in [n]$, we have
\begin{align*}
    | (A {\bf 1}_n)_i - (B {\bf 1}_n)_i | \leq  2 n \eta 
    \leq  \eta \cdot | ( A {\bf 1}_n )_i |
\end{align*}
where the first step follows from Eq.\eqref{eq:A_1_B_1} and the second step is from Eq.\eqref{eq:A_1_n}.
\end{proof}


\begin{proof}[Proof of Lemma~\ref{lem:D(A)_A}]
 We proof each part below.
\paragraph{Proof of Part 1.}
    We have 
    \begin{align*}
        |D(A)_{i,i} - D(B)_{i,i}| \leq & ~ | (A {\bf 1}_n)_i - (B {\bf 1}_n)_i | \\
        \leq & ~ \eta \cdot | ( A {\bf 1}_n )_i |\\ 
         \leq & ~ \eta \cdot |D(A)|_{i,i}\\ 
    \end{align*}
where the first step is from Definition~\ref{def:attention_matrix}, the second step is based on {\bf Part 5} of Lemma~\ref{lem:perturb_tools} and the third step is based on Definition~\ref{def:attention_matrix}.

\paragraph{Proof of Part 2.}
The error analysis for this part follows a similar approach as in {\bf Part 1}. Due to its similarity, we will omit the details here.
\paragraph{Proof of Part 3.}
Based on {\bf Part 1.} of Lemma~\ref{lem:perturb_tools}, for $j \notin S_i$, we have $| A_{i,j} - B_{i,j} | \leq 2 \eta$.
Based on {\bf Part 2.} of Lemma~\ref{lem:perturb_tools}, for $j \in S_i$, $| A_{i,j} - B_{i,j}|= 0$.
Furthermore, taking into account the aforementioned findings, we can conclude that $| A_{i,j}-B_{i,j}| \leq 2 \eta$.
\end{proof}

\begin{proof}[Proof of Lemma~\ref{lem:error_analysis:A}]
  We first decompose the difference into
    \begin{align}\label{eq:f1_f2}
        & ~ \|D( A )^{-1} A - D( B )^{-1} B \|_\infty \notag \\
    \le & ~ \|D( A )^{-1}A - D( B )^{-1} A \|_\infty + \|D( B )^{-1} A - D( B )^{-1} B \|_\infty  \notag \\
    = & ~ F_1 + F_2.
    \end{align}
Now, we will provide the upper bounds for $F_1$ and $F_2$, respectively.
    We have 
    \begin{align}\label{eq:f1}
        F_1= & ~ \|D( A )^{-1}A - D( B )^{-1} A \|_\infty \notag \\
        = & ~ \max_{i \in [n], j \in [n]} \{| (D( A )^{-1}A - D( B )^{-1} A)_{i,j}|\} \notag \\
        = & ~ \max_{i \in [n], j \in [n]} \{| A_{i,j}(D( A )^{-1}_{i,i} - D( B )^{-1}_{i,i})|\} \notag \\
        \leq & ~ \max_{i \in [n], j \in [n]} \{| A_{i,j} | \cdot | \frac{D(A)_{i,i}-D(B)_{i,i}}{D(A)_{i,i}D(B)_{i,i}} |\} \notag \\
        \leq & ~  \max_{i \in [n], j \in [n]} \{| \frac{\eta D(B)_{i,i}}{D(A)_{i,i}D(B)_{i,i}} | \cdot |  A_{i,j} |\} \notag \\
    = & ~ \eta \cdot \max_{i\in[n], j \in [n]}\{|D(A)^{-1}_{i,i} | \cdot |  A_{i,j}| \}\notag \\
        \leq & ~ \eta
    \end{align}
where the first step follows from the definition of $F_1$, the second step is also based on the definition of infinity norm, the third step is due to simple algebra, the fourth step comes from triangle inequality, the fifth step is because of {\bf Part 1.} of Theorem~\ref{lem:D(A)_A}, the sixth step is due to simple algebra and the last step is due to Definition~\ref{def:attention_matrix}.

    We have 
    \begin{align} \label{eq:f2}
         F_2 = & ~ \|D( B )^{-1}A - D( B )^{-1} B \|_\infty \notag \\
        = & ~ \max_{i \in [n], j \in [n]} \{| (D( B )^{-1}A - D( B )^{-1} B)_{i,j}|\} \notag \\
        = & ~ \max_{i \in [n], j \in [n]} \{| D(B)^{-1}_{i,i}| \cdot |A - B|_{i,j}\} \notag \\
        \leq & ~ 2 \eta  \cdot \max_{i \in [n], j \in [n]} \{|D(B)_{i,i}^{-1}|\} \notag \\
        \leq & ~ 2 \eta 
    \end{align}
where the first step comes from the definition, the second step is because of the definition of infinity form, the third step follows from simple algebra, and the last step is from {\bf Part 4} of Lemma~\ref{lem:perturb_tools}.

By combining the aforementioned findings and conclusions, we can establish the following result.
\begin{align*}
      & ~ \|D( A )^{-1} A - D( B )^{-1} B \|_\infty \\
    = & ~ F_1 + F_2 \\
    = & ~ O(\eta)
\end{align*}
where the first step follows from Eq.~\eqref{eq:f1_f2} and the second step follows from Eq.~\eqref{eq:f1} and Eq.~\eqref{eq:f2}.
\end{proof}
\section{Omitted Proofs in Section~\ref{sec:analysis_of_attention_computation_final}}\label{sec:omit_proof_4}

\begin{proof}[Proof of Lemma~\ref{lem:sparsity_B_quantum}]

{\bf Proof of Part 1.}
For $i \in [n]$, we will focus on a vector $(Q K^\top)_{i,*}$. 

Given that $j \in [d]$, we define $u(j)$ such that
\begin{itemize}
    \item $u(j) = 1$ if $(Q K^\top)_{i,j} > \tau$
    \item $u(j) = 0$ else.
\end{itemize}

According to Part 2 of Theorem~\ref{thm:grover_search}, we can use quantum algorithms to efficiently locate all elements $j \in [n]$ for which $u(j) = 1$.

Give that we compute $u(j)$ in $O(d)$, the time complexity of the quantum algorithm is $\wt{O}(\sqrt{n k} d )$.

{\bf Proof of Part 2.}

Based on the proof above, we will output a sparse vector $B_{i,*}$ here. The time complexity of compute the sparse vector can be divided into two parts. One is to find the satisfied element in $\wt{O}(\sqrt{n k} d)$, which has been proven above. 

Given that $k$ represents the upper bound on the number of satisfied elements, the matrix computation specifically targets those satisfied elements, resulting in a time complexity of $O(kd)$.

The proof is now complete.
\end{proof}

\begin{proof}[Proof of Lemma~\ref{lem:error_analysis}]
We have 
\begin{align*}
    &\| D(A)^{-1} A V - D(B)^{-1} B V \|_\infty \\
    \leq &~ \| D(A)^{-1} B V - D(B)^{-1} B V \|_\infty \\
    &+ \| D(A)^{-1} B V - D(A)^{-1} A V \|_\infty.
\end{align*}
For each $(i,j) \in [n] \times [d]$,
    Based on Lemma~\ref{lem:error_analysis:A}, we have
    \begin{align}\label{eq:D_B_V}
        &| (D(A)^{-1} B V - D(B)^{-1} B V)_{i,j} | \notag \\
        = & ~ |\sum_{l = 1}^n  (D(B)^{-1}_{i,i} - D(A)^{-1}_{i,i}) \cdot B_{i,l} \cdot V_{l,j}| \notag \\
        \leq & ~ \sum_{l = 1}^n  |(D(B)^{-1}_{i,i} - D(A)^{-1}_{i,i}) \cdot B_{i,l}| \cdot  \| V \|_\infty \notag \\
        \leq & ~ \sum_{l = 1}^n |\frac{D(B)_{i,i} - D(A)_{i,i}}{D(B)_{i,i}D(A)_{i,i}} B_{i,l}| \cdot \|V\|_\infty \notag \\
        \leq & ~ \eta \cdot \sum_{l=1}^n | D(B)_{i}^{-1} B_{i,l}| \cdot \| V \|_\infty \notag \\
        = & ~ \eta \cdot | \sum_{l=1}^n D(B)_{i}^{-1} B_{i,l} | \cdot \| V \|_\infty \notag \\
        = & ~ \eta \cdot \| V \|_\infty \notag \\
        = & ~ O(\eta^2)
    \end{align}
where the first step follows from simple algebra, the second step is based on the definition of infinity norm, the third step is because of simple algebra, the forth step is from Lemma~\ref{lem:D(A)_A}, the fifth step is because of Definition~\ref{def:attention_matrix}, the sixth step is based on Definition~\ref{def:attention_matrix}, and the last step is because of $\| V \|_\infty \leq \eta$.

For each $( i,j ) \in [n] \times [d]$, 
we have 
\begin{align}\label{eq:D_A_V}
     | (D(A)^{-1} B V - D(A)^{-1} A V)_{i,j} | = & ~ |\sum_{l = 1}^n  (D(A)^{-1}_{i,i} ( B_{i,l} - A_{i,l}) \cdot V_{l,j}| \notag \\
     \leq & ~ \sum_{l = 1}^n  |(D(A)^{-1}_{i,i}| \cdot |( B_{i,l} - A_{i,l})| \cdot \| V \|_\infty \notag \\
     = & ~ O(\eta^2)
\end{align}
where the first step is based on simple algebra, the second step is because of triangle inequality, and the last step is based on $\| V \|_\infty$.

The proof is enhanced by combining  Eq.~\eqref{eq:D_B_V} and Eq.~\eqref{eq:D_A_V}.
\end{proof}

\end{document}